\documentclass[graybox]{svmult}

\usepackage{type1cm}        % activate if the above 3 fonts are
\usepackage{makeidx}         % allows index generation
\usepackage{graphicx}        % standard LaTeX graphics tool
\usepackage{multicol}        % used for the two-column index
\usepackage[bottom]{footmisc}% places footnotes at page bottom

\usepackage{newtxtext}       % 
\usepackage[varvw]{newtxmath}       % selects Times Roman as basic font

\newcommand{\pa}{\partial}

\newcommand{\bx}{\mathbf{x}}

\newcommand{\nn}{\nonumber}

\makeindex             % used for the subject index
\begin{document}

\title*{A reaction-diffusion model for relapsing-remitting multiple sclerosis with a treatment term}
 \titlerunning{A reaction-diffusion model for relapsing-remitting multiple sclerosis} 
\author{Romina Travaglini\orcidID{0000-0003-4107-1764}}
\institute{Romina Travaglini \at Istituto Nazionale di Alta Matematica, Università degli Studi di Parma \\ \email{romina.travaglini@unipr.it}}
\maketitle

\abstract*{We present a mathematical study for the development of multiple sclerosis based on a reaction-diffusion system. The model describes interactions among different populations of human cells, motion of immune cells stimulated by cytokines, consumption of myelin sheath due to anomalously activated lymphocytes and its restoration by oligodendrocytes. Successively, we introduce a therapy term representing injection of low-dose IL-2 interleukine. A natural step is then to  study the system, investigating the formation of spatial patterns by means of a Turing instability analysis of the problem. In particular, we get spatial patterns oscillating in time that may reproduce brain lesions characteristic of the early stage of the pathology, in both non-treatment and treatment scenarios.}

\abstract{We present a mathematical study for the development of multiple sclerosis based on a reaction-diffusion system. The model describes interactions among different populations of human cells, motion of immune cells stimulated by cytokines, consumption of myelin sheath due to anomalously activated lymphocytes and its restoration by oligodendrocytes. Successively, we introduce a therapy term representing injection of low-dose IL-2 interleukine. A natural step is then to  study the system, investigating the formation of spatial patterns by means of a Turing instability analysis of the problem. In particular, we get spatial patterns oscillating in time that may reproduce brain lesions characteristic of the early stage of the pathology, in both non-treatment and treatment scenarios.}

\section{Introduction}
\label{sec:1}

Multiple Sclerosis (MS) is one of the most severe autoimmune disorders. Its primary pathological characteristic is injury of the myelin sheath, that  surrounds axons in the nervous system (CNS) and favors the transmission of cerebral impulses. Such lesions can be observed,  through MRI, as development of focal plaques in the white matter.
	
It is widely accepted that the biological dynamics behind MS involves immune cells as T-cells, B-cells, macrophages and microglia. These cells may be dysfunctionally activated against cells producing myelin or myelin itself. This activation may induce an autoimmune cascade, promoted by production of proinflammatory cytokines, i.e. molecules that attract and stimulate clonal expansion of immune cells.

Medical literature on MS reports a wide range of cases in terms of clinical course, features of lesions and associated irreversible neurological symptoms. Usually, the early phase of the disease is characterized by acute inflammation and formation of active lesions. At the same time,  also a restoration process, referred as remyelination, takes place, resulting in the formation of "shadow" plaques. This demyelination-remyelination process may continue to alternate for months or even years and this phase is indicated as
 relapsing-remitting MS (RRMS). This stage is usually followed by a second phase, called secondary progressive MS (SPMS), when restoration of active myelin lesions is less frequent, resulting in a progressive neurodegeneration. Further details about medical description of MS may be found in \cite{lassmann2005multiple,lassmann2007immunopathology,lassmann2012progressive,mahad2015pathological} and references therein.

 For our purposes, it must be underlined the fact  that self-reactive immune cells can be found  in non-pathological conditions as well \cite{danke2004autoreactive}. But in this case, the action of specific cells called immunosuppressors can  inhibit or kill cells presenting the antigen and activated immune cells. In autoimmune conditions as MS, though, both the number and the efficiency of these natural killers are lacking  \cite{hoglund2013one,mimpen2020natural,zozulya2008role}.

%-Modelli esistenti macroscopici
In a recent work  \cite{travaglini2023reaction} a mathematical model describing peculiar dynamics of MS has been proposed. In particular, starting from kinetic models of active particles applied to autoimmune diseases as  \cite{delitala2013mathematical,kolev2018mathematical}, followed by  \cite{ramos2019kinetic,della2022mathematical}, authors derive a system of partial-differential equations of reaction-diffusion type with a chemotaxis term obtained by a diffusive limit in aproper time scaling. Populations of cells and biological substances there considered are: antigen presenting cells (APCs), self-reactive leukocytes (SRLs), immunosuppressive cells (ISCs), cytokines and consumed myelin.
Through a Turing instability analysis, they reproduce formation of one-dimensional  patterns whose dynamics is comparable {to}  RRMS or SPMS.

{The aim of the present work is to focus on the RRMS dynamics, introducing a treatment term representing injection of low-dose interleukin L-2 (IL-2). As reported in literature, in fact, in autoimmune conditions \cite{furtado2002interleukin}, low-dose IL-2 may stimulate expansion of ISCs  \cite{klatzmann2015promise} and enhance their suppressive function. The manner in which these effects can be achieved in the specific case of MS, coupled with the stimulation of ISC capacity to promote remyelination, is currently under investigation \cite{louapre2023randomized}.}

 {In this paper, we perform a reduction of the reaction-diffusion system derived in \cite{travaglini2023reaction} and introduce the new term representing effects of therapy. Then we propose a Turing instability analysis of the system, in order to find conditions on parameters leading to formation and restoration of plaques. Results will be validated and compared by means of numerical simulations in a two-dimensional domain.}

\section{The model}
\label{sec:2}
We recall here the macroscopic reaction-diffusion model derived in  \cite{travaglini2023reaction} that describes formation and restoration of myelin plaques in multiple sclerosis, in a non-dimensional form. The system of reaction-diffusion equations reads as
\begin{eqnarray}\nn
	\label{eq:rdsA} \frac{\pa A}{\pa t}&=\,&1+\beta\,A\,R - A\,S - \zeta A , \\[1mm]\nn
	\label{eq:rdsS1} \frac{\pa S}{\pa t}&=\,& \mu\,A\,S - \,S , \\[1mm]\nn
	 \frac{\pa R}{\pa t}&=\,&\nabla_{\bx}\cdot\left(\Phi_0(R)\,\nabla_{\bx}\, R-{\chi} \,\Phi_1(R) R\,\nabla_{\bx}\,C\right)+\eta\,A\,R - \psi\,R\,S - \theta\,R , \\[1mm]\nn
	\frac{\pa C}{\pa t}&=\,&\delta \Delta_{\bx}\,C+A\, R-\tau\,C,\\[1mm]\nn
	\label{eq:rdsE1} \frac{\pa E}{\pa t}&=\,& \frac{\gamma\,R}{\xi+R}\,R\,(1-E)-\lambda\,E,
\end{eqnarray}
where involved quantities are macroscopic densities for
\begin{itemize}
	\item[] {\large $A$} -- Self-antigen presenting cells (SAPCs).
	\item[] {\large $S$} -- Immunosuppressive cells (ISCs).
	\item[] {\large $R$} -- Self-reactive leukocytes (SRLs).
	\item[] {\large $C$} -- Cytokines.
	\item[] {\large $E $} -- Destroyed myelin,
\end{itemize}
depending on $t\in\mathbb{R}_0^+$ and $\bx\in\Omega$, with $\Omega$ bounded domain in $\mathbb{R}^2$. Moreover, we choose $$\Phi_0(y)=\cos\left(\frac\pi2\,y\right)+\frac\pi2\,y\,\sin\left(\frac\pi2\,y\right), \quad\Phi_1(y)=\cos\left(\frac\pi2\,y\right).$$ A detailed discussion about more general choice for spatial domain and functions can be found in \cite{travaglini2023reaction}.

As first, since the aim of the present work is to focus on the effects of low-dose IL-2 on ISCs, SRLs and myelin sheath, we assume that the population of SAPCs constitutes a background which SRLs and ISCs interact with. Henceforth, we suppose that SAPCs density is constant at equilibrium, i.e.
\begin{eqnarray}\nn
	A&=\,&\frac{1}{\zeta -\beta\,R+S}.
\end{eqnarray}
 Moreover, we introduce the parameter $\alpha\in[0,1)$ representing low-dose IL-2 delivery, obtaining the system

\begin{eqnarray}
	\label{eq:S} \frac{\pa S}{\pa t}&=&\,S\, \left(\frac{\mu }{\zeta -\beta\,R+S}-\Gamma\right), \\[1mm]\nn
	\label{eq:rdsR} \frac{\pa R}{\pa t}&=\,&\nabla_{\bx}\cdot\left(\Phi_0(R)\,\nabla_{\bx}\, R-{\xi} \,\Phi_1(R) R\,\nabla_{\bx}\,C\right) \\[1mm]
	&&\,+
	R \left( \frac{\eta }{\zeta -\beta\,R+S}-  \Psi \,S-\theta\,\right), \\[1mm]
	\label{eq:rdsC}\frac{\pa C}{\pa t}&=\,&\delta \Delta_{\bx}\,C+\frac{R}{\zeta -\beta\,R+S}-\tau\,C,\\[1mm]
	\label{eq:E} \frac{\pa E}{\pa t}&=\,& \frac{\gamma\,R}{\xi+R}\,R\,(1-E)-\Lambda\,E,
\end{eqnarray}
where $\Gamma=(1-\alpha)$, $\Psi=\psi\,(1+\alpha)$, $\Lambda=\lambda\,(1+\alpha)$.

\subsection{Turing instability}
\label{subsec:2.2}
We study macroscopic system \eqref{eq:S}-\eqref{eq:E} by means of Turing instability analysis. More in detail, we aim at finding a particular range for parameters leading to the appearance of oscillating spatial patterns that may reproduce the {appearance} (and possible reconstruction) of myelin plaques.

		We consider system \eqref{eq:S}-\eqref{eq:E} taking as initial data 
	\begin{equation}\nn\label{InCond2}\nn
		\mathbf{U}(0,\bx)=\mathbf{U}_0(\bx)\geq 0, \mbox{ with } R_0(\bx)\leq 1,\,E_0<1,
	\end{equation}
	and imposing zer{o}-flux at the boundary
	\begin{equation}\nn\label{ZeroFlux}
	\Big(\Phi_0(R)\,\nabla_{\bx}\, R-{\xi} \,\Phi_1(R) R\,\nabla_{\bx}\,C \Big)\cdot {\bf \widehat n}=0,\quad \nabla_{\bx} C\cdot {\bf \widehat n}=0,
	\end{equation}
	with  ${\bf \widehat n}$ {being} the external unit normal to {the boundary} $\pa\Omega$.
	Turing instability \cite{turing52} occurs when a spatially homogeneous steady state  for the system  without spatial gradients turns to be unstable when diffusive and chemotactic terms are added. 
To this aim, we individuate an equilibrium 
	$U_1= (S_1,\,R_1,\,C_1,\,E_1)$ for the system \eqref{eq:S}-\eqref{eq:E}, in spatially homogeneous conditions, that is biologically relevant, i.e. that belongs to the set $
	\mathcal E=\left\{  0< R(t,\bx)\leq 1,\, C(t,\bx)> 0,\, 0<E(t,\bx)<1\right\}. $
	Thus we find
	\begin{equation}\nn\label{U}
	U_1 = \,\left(\frac{\Gamma\,\eta -\theta\,\mu }{\mu \,\Psi },\,\frac{\Gamma\,\zeta\,\mu\,\Psi +\Gamma^2\,\eta -\Gamma\,\theta\,\mu -\mu ^2\,\Psi}{\beta\,\mu\,\Psi },\,       
	\frac{1}{\mu\tau}\,R_1,\,\frac{R_1^2 \gamma}{R_1^2\, \gamma + 
		R_1\, \Lambda + \Lambda\, \xi} \right),
	\end{equation}
    that belongs to the set $\mathcal E$ when the following conditions on parameters are satisfied 
	\begin{equation}\label{EqPos0}
	\theta<\frac{\Gamma\,\eta}{\mu},\quad\bar\theta > \theta > \bar\theta-{\beta\Psi},
	\quad\mbox{with}\quad
	\bar\theta:=\frac{\Gamma\,\eta}{\mu}+\frac{\Psi(\Gamma\,\zeta-\mu)}{\Gamma}.
	\end{equation}
		We find it convenient to take $\Gamma\,\zeta>\mu$, in such a way conditions in \eqref{EqPos0} become
			\begin{equation}\nn\label{EqPos}
			\frac{\Gamma\,\eta}{\mu} > \theta > \bar\theta-{\beta\Psi}.
		\end{equation}
Linearizing system \eqref{eq:S}-\eqref{eq:E} around equilibrium $U_1$, we have
{ t}he Jacobian matrix defined as
\begin{equation} \nn
	\mathbb A=
\left(
\begin{array}{cccc}
	-\frac{\Gamma^2\,S_1}{\mu } & \frac{\Gamma ^2 \,\beta\, \,S_1}{\mu } & 0 & 0 \\ \\
	-\frac{R_1 \left(\Gamma^2\eta +\mu ^2 \Psi \right)}{\mu ^2} & \frac{\Gamma^2\,\beta \,\eta\,  R_1}{\mu ^2} & 0 & 0 \\ \\
	-\frac{\Gamma^2\,R_1}{\mu ^2} & \frac{\Gamma }{\mu }+\frac{\Gamma ^2 \,\beta\,{R_1}}{\mu ^2} & -\tau  & 0 \\ \\
	0 & \frac{\gamma\,  \Lambda\,\  R_1 (2\,\xi +R_1)}{(\xi +R_1) (\xi\,\Lambda +R_1 (\Lambda +\gamma\,R_1))} & 0 &  -\frac{\gamma  \,R_1^2}{\xi +R_1} -\Lambda
\end{array}
\right).
\end{equation}
{The e}igenvalues of the matrix $\mathbb A$ are \ $-\tau$ and $-\Lambda - \frac{\gamma  R_1^2}{\xi +R_1} $, 
and the eigenvalues $\sigma_1,\,\sigma_2$ of the 2x2 square submatrix in the top left corner. It holds that 
\begin{equation}\nn
 \sigma_1+\sigma_2=\frac{\Gamma^2}{\mu}\left(\frac{R_1\,\eta\,\beta}{\mu}-S_1\right),\quad \sigma_1\,\sigma_2=\frac{\Gamma^2\,\beta\,\Psi}{\mu}\,R_1\,S_1>0. 
\end{equation}
Thus,  the equilibrium $U_1$ is locally asymptotically stable holding the condition 
\begin{equation}\nn
\theta<\tilde\theta, \quad \tilde\theta=\frac{\Gamma\,\eta }{\mu }+\frac{\eta\,  \Psi\,  (\Gamma\,\zeta -\mu )}{\Gamma\,(\eta-\mu ) }.
\end{equation} 

Now, we consider the complete system with the spatial derivative terms. When linearizing, we get the diffusion matrix $\mathbb D$ having all entries null except for elements in positions (2,2), (2,3) and (3,3), that are equal to $\Phi_0(R_1)$, $ -\Lambda \Phi_1(R_1) R_1$ and $\delta$, respectively.
%Solutions of system \eqref{SistW} written in Fourier series are of type
%\begin{equation} \label{Wgen} {\bf W}({\bf x},t)=\sum_{k}c_ke^{\xi_k t}\,{{{\bf \overline W}_k}({\bf x})}\,,\end{equation}
%where {the} wavenumbers $k \in \mathbb{N}$ and the eigenfunctions $\overline{\bf W}_k({\bf x})$ {define} the solution of the time--independent problem.
% \begin{equation}\left\{\begin{array}{ll}\Delta_{\bf x}\tilde{\bf W}+k^2\bar{\bf W}={\bf 0}&\mbox{ on }(0,\infty)\times\Gamma_{\bx}\\&\\
%{\bf \hat n}\cdot \nabla_{\bf x}\bar{\bf W}=0 &\mbox{ on }(0,\infty)\times\pa\Gamma_{\bx}.\end{array}\right.
%\end{equation}
Formation of spatial pattern may arise when it is possible to find an interval $(k_1,k_2)$ such that for any $k_1\leq k\leq k_2$,  $Re(\lambda_{ k})>0$, being $\lambda_k$ an eigenvalue for the matrix $\mathbb A-k^2\mathbb D$. We can individuate only a sufficient condition to have eigenvalues with positive real part, that is {$det(\mathbb A-k^2\mathbb D)<0$}, where
 \begin{equation}\nn
\det({\mathbb A} - k^2 {{\mathbb D}}) =\frac{\Gamma^2\,S_1}{\mu } \left(\Lambda +\frac{\gamma  R_1^2}{\xi +R_1}\right)\,h(k^2),
\end{equation}
 \begin{equation}\nn
h(k^2):= k^4 \delta\,\Phi_0(R_1) + k^2 q + \tau\,R_1\,\beta\,\Psi,
\end{equation}
 \begin{equation}\nn
q= R_1\,\delta\,\beta\,\Psi  - \frac{\Gamma\,\chi}{\mu}\,\Phi_1(R_1)\,R_1 + \Phi_0(R_1)\,\tau.
\end{equation}
It is straightforward to observe that the condition $\det({\mathbb A} - k^2 {{\mathbb D}})< 0$
is satisfied if and only if we have the following two
\begin{equation} \nn
\label{TurCond}
q<0\mbox{ and }q^2-4\,R_1\,\delta\,\beta\,\Psi \,\Phi_0(R_1)\tau>0,
\end{equation}
that, after computations, lead to the a threshold value for $\chi$ that reads as
\begin{equation}\nn
\chi> \mu\,\frac{2\sqrt{ \Phi_0(R_1)\,R_1\,\delta\, \tau\,\beta\,\Psi}+\delta\,R_1\,\beta\,\Psi +\Phi_0(R_1)\,\tau}{\Gamma\,\Phi_1(R_1)\, R_1}.
\label{eq:pf} 
\end{equation}

\section{Numerical simulations}
\label{sec:3}
Analytical results obtained in the previous section are now applied to the study of quantities involved in system \eqref{eq:S}-\eqref{eq:E} in space and time. We perform numerical simulation using \texttt{VisualPDE} \cite{walker2023visualpde} in a square domain of size $L=10$. In particular, we focus on the dynamics of destroyed myelin $E$. We fix the following parameters 
\begin{equation}
\begin{array}{c}
\beta = 0.8,\quad \zeta = 2.05,\quad \mu = 2,\quad \tau = 0.5,\quad \eta = 1,\quad \delta=0.1,\quad \theta=0.4,\\\\ \psi=0.3,\quad\gamma=2,\quad\xi=1,\quad\lambda=0.3.\label{Pars}
\end{array}
\end{equation}

\begin{figure}[h]
	\includegraphics[scale=.8]{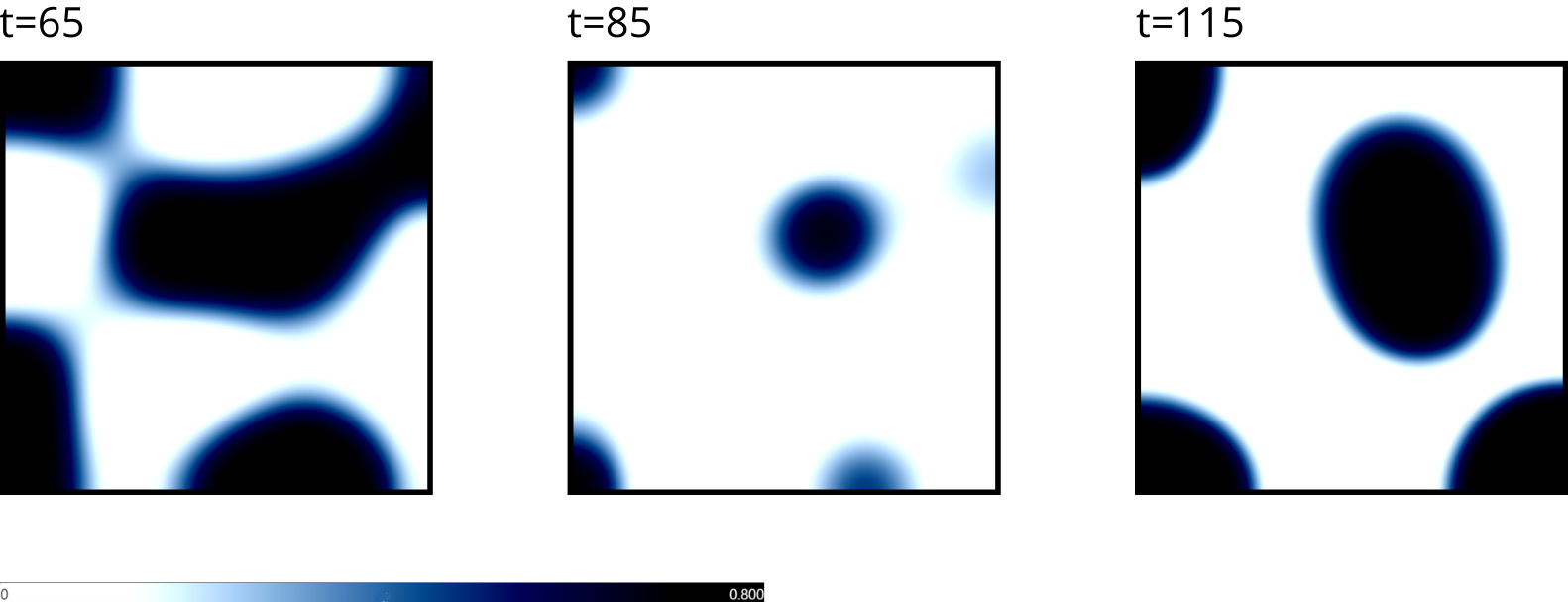}
	\caption{Evolution of quantity $E$ (destroyed myelin), taking values as in \eqref{Pars}, $\alpha=0$, $\chi=7$.}
	\label{Fig1}       % Give a unique label
\end{figure}

Vales above individuate an interval for the treatment parameter $\alpha$ that is  $0\leq\alpha<0.095$.
We start by simulating the non-treatment, setting  $\alpha=0$ and as initial conditions a perturbation of equilibrium $U_1$ for quantities $S,R,C$ while $E$ is taken uniformly null at the beginning. In this case we have a critical  $\chi_c\approx 4.59$, then we fix $\chi=7$ and we obtain results shown in Figure \ref{Fig1}. It can be observed that the myelin plaques (areas where the density of $E$ is higher) have an oscillatory behavior in time, as happens in the case of RRMS. 

\begin{figure}[h]
	\includegraphics[scale=.8]{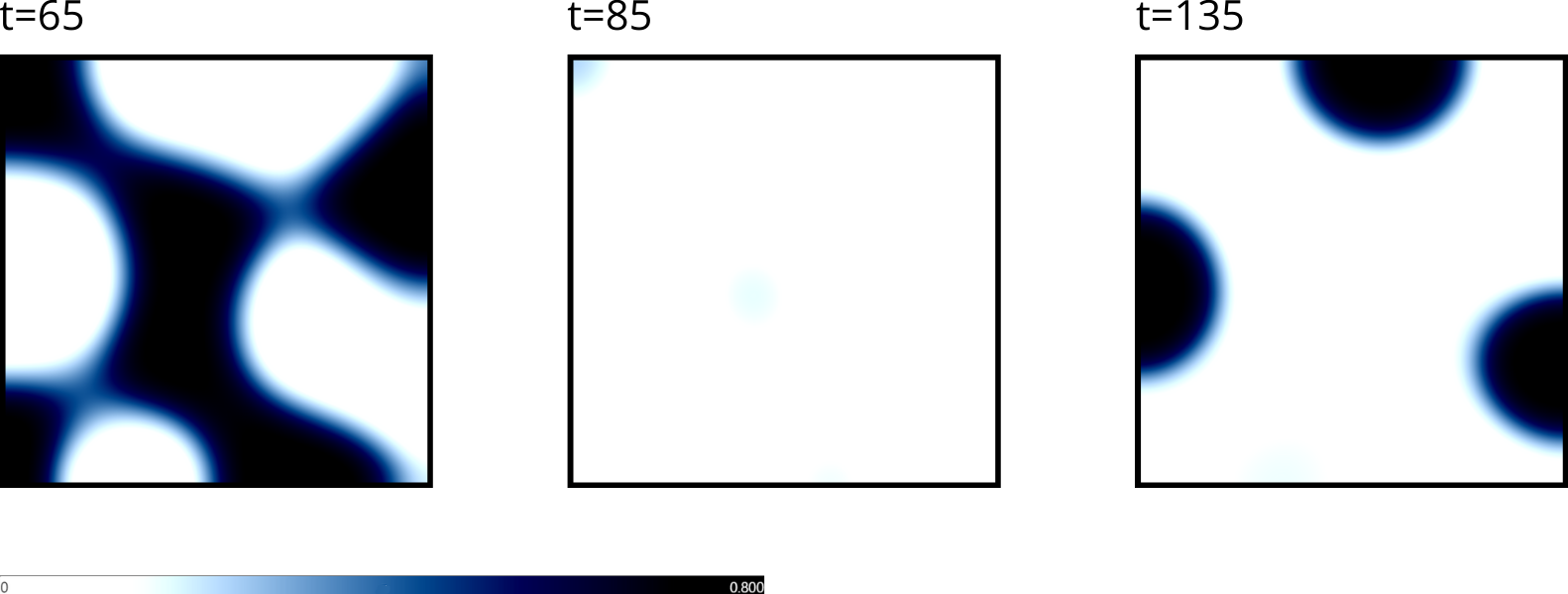}
	\caption{Evolution of quantity $E$ (destroyed myelin), taking values as in \eqref{Pars}, $\alpha=0$ for $0<t<65$ and  $\alpha=0.05$ for $t>65$, $\chi=7$.}
	\label{Fig2}       % Give a unique label
\end{figure}

For the second case, instead, we take $\alpha=0$ and initial conditions as in the previous case, then, when lesions appear,  we increase $\alpha=0.05$. The critical value for $\chi$ now is $\chi_c\approx 6.37$, hence we keep $\chi=7$, obtaining dynamics reported in Figure \ref{Fig2}. It is possible to observe that, at time $t=85$ plaques completely disappear, differently from the first case. Moreover, lesions take a longer time to form again, we underline that this effect is indirectly induced by a lower decreasing and a major suppressive efficiency of ISCs, given by parameters $\Lambda$ and $\Psi$, respectively. This fact affects negatively the growth of SRLs and, consequently the consumption of sane myelin.

\section{Conclusions}
\label{sec:4}
In this work, we considered the model proposed in \cite{travaglini2023reaction}, where a reaction-diffusion model for the study of multiple sclerosis is derived from kinetic level and analyzed. In particular, we focused on results that may describe the first phase of multiple sclerosis, called relapsing-remitting phase, characterized by formation and restoration of plaques in the white matter of the central nervous system. Our aim was, on the one hand to reproduce the behavior of the model performing simulations on a two-dimensional domain, on the other hand to study the effects of possible treatment terms. For this reason, we performed a reduction of the original system, introducing a term representing injection of low-dose interleukin L-2 and we went through a Turing instability analysis. We found conditions on parameters allowing for formation of patterns and we validated results numerically. We reproduced the periodic formation and restoration of myelin plaques, confirming that the treatment term helps the remyelination and delays demyelination.

\begin{acknowledgement}
Tha author is a post-doc fellow of the National Institute of Advanced Mathematics (INdAM), Italy. The research was carried out in the frame of activities sponsored by the Cost Action CA18232.
\end{acknowledgement}
\ethics{Competing Interests}{
The research work was supported by INdAM (National Institute of Advanced Mathematics), by the Portuguese national funds (OE), through FCT/MCTES FCT/MCTES  (Fundação para a Ciência e a Tecnologia) Projects  UIDB/00013/2020,  UIDP/00013/2020, \\ PTDC/03091/2022 (“Mathematical Modelling of Multi-scale Control Systems: applications to human diseases (CoSysM3)”), and by University of Parma through the action Bando di Ateneo 2022 per la ricerca co-funded by MUR-Italian Ministry of Universities and Research - D.M. 737/2021 - PNR - PNRR - NextGenerationEU” (project: "Collective and self-organised dynamics: kinetic and network approaches")}

\eject

\bibliographystyle{abbrv}
\bibliography{biblioAbb}

\end{document}